# Microphysics of Water Clouds in the Atmospheres of Y Dwarfs and Temperate Giant Planets

James Mang[1] , Peter Gao[2] , Callie E. Hood[3] , Jonathan J. Fortney[3] , Natasha Batalha[4] , Xinting Yu[5] , and Imke de Pater[6]

[1] Department of Astronomy, University of Texas at Austin, Austin, TX 78712, USA; j_mang@utexas.edu
[2] Earth & Planets Laboratory, Carnegie Institution for Science, 5241 Broad Branch Rd NW, Washington, DC 20015, USA
[3] Department of Astronomy and Astrophysics, University of California, Santa Cruz, CA 95064, USA
[4] NASA Ames Research Center Moffett Field, CA 94035, USA
[5] Department of Earth and Planetary Sciences, University of California, Santa Cruz, CA 95064, USA
[6] Department of Astronomy, University of California, Berkeley, CA 94720, USA
*Received 2021 October 13; revised 2022 January 31; accepted 2022 February 2; published 2022 March 15*

## Abstract

Water clouds are expected to form on Y dwarfs and giant planets with equilibrium temperatures near or below that of Earth, drastically altering their atmospheric compositions and their albedos and thermal emission spectra. Here we use the 1D Community Aerosol and Radiation Model for Atmospheres (CARMA) to investigate the microphysics of water clouds on cool substellar worlds to constrain their typical particle sizes and vertical extent, taking into consideration nucleation and condensation, which have not been considered in detail for water clouds in H/He atmospheres. We compute a small grid of Y-dwarf and temperate giant-exoplanet atmosphere models with water clouds forming through homogeneous nucleation and heterogeneous nucleation on cloud condensation nuclei composed of meteoritic dust, organic photochemical hazes, and upwelled potassium chloride cloud particles. We present comparisons with the Ackerman & Marley parameterization of cloud physics to extract the optimal sedimentation efficiency parameter ($f_{sed}$) using Virga. We find that no Virga model replicates the CARMA water clouds exactly and that a transition in $f_{sed}$ occurs from the base of the cloud to the cloud top. Furthermore, we generate simulated thermal emission and geometric albedo spectra and find large, wavelength-dependent differences between the CARMA and Virga models, with different gas absorption bands reacting differently to the different cloud distributions and particularly large differences in the $M$ band. Therefore, constraining the vertically dependent properties of water clouds will be essential to estimate the gas abundances in these atmospheres.

*Unified Astronomy Thesaurus concepts:* Y dwarfs (1827); Brown dwarfs (185); Extrasolar gaseous giant planets (509); Exoplanet atmospheres (487); Exoplanet atmospheric composition (2021); Atmospheric clouds (2180); Planetary atmospheres (1244)

## 1. Introduction

The discovery of hot giant exoplanets drastically diversified our picture of giant-planet formation and composition. Since then, our efforts have led to the detection of even cooler giant exoplanets, closing the gap between exoplanets and the giant planets in our solar system. As we approach temperatures below ~400 K, highly reflective water clouds are predicted to form, which should significantly alter the atmosphere due to the exchange of water vapor opacity for water-cloud opacity, the release of latent heat, and the increase in planetary geometric albedo (e.g., Marley et al. 1999; Burrows et al. 2003; Morley et al. 2014a; Tang et al. 2021). These objects would stand apart from both hotter planets that are devoid of water clouds, as well as the solar system giant planets that are cold enough for the formation of, e.g., ammonia- and methane-ice clouds atop the water clouds. Thus, studying the atmospheres of such temperate worlds will help us probe a unique part of planetary parameter space.

A growing number of exoplanets have now been discovered that could allow for water-cloud formation, such as K2-18b (Benneke et al. 2019; Tsiaras et al. 2019), LHS1140b (Dittmann et al. 2017; Edwards et al. 2021), Kepler-51d

(Masuda 2014; Libby-Roberts et al. 2020), a giant-planet candidate transiting the white dwarf WD 1856+534 (TIC 267574918; Vanderburg et al. 2020), PH-2b and Kepler-103 c (Dubber et al. 2019), and Kepler-167e (Dalba & Tamburo 2019). Additionally, water clouds could form on Y-type brown dwarfs, particularly those with effective temperature $T_{eff} < 450$ K. Y dwarfs represent a transition between low-mass stars and giant planets and allow for studies of planetary atmosphere properties at higher gravities (Cushing et al. 2011). Observations of the coolest Y dwarfs suggest the existence of water-ice clouds (Faherty et al. 2014; Leggett et al. 2015; Skemer et al. 2016; Morley et al. 2018), though it is difficult to reproduce their near-IR spectral energy distributions with existing cloudy models (Luhman & Esplin 2016). Furthermore, with the launch of the James Webb (JWST) and Nancy Grace Roman Space Telescopes, observations of thermal emission and reflected light from cool objects possessing water clouds will multiply (Morley et al. 2014b; Lupu et al. 2016; Lacy et al. 2019), motivating the need for more rigorous studies of these worlds.

Unlike investigations of water clouds in terrestrial exoplanet atmospheres (e.g., Zsom et al. 2012; Yang & Abbot 2014; Komacek et al. 2020; Charnay et al. 2021), there have been few modeling studies of water clouds in H/He atmospheres. These have included cloud models of various complexity, from those that are highly parameterized (Lupu et al. 2016) to those that take into account saturation, sedimentation, and atmospheric mixing in calculating cloud particle-size and spatial distributions (e.g., Marley et al. 1999; Ackerman & Marley 2001; Burrows et al. 2003; Morley et al. 2014b; MacDonald et al. 2018; Hu 2019). A recent study also investigated the effects of water latent heat release on the thermal structure of brown dwarfs (Tang et al. 2021). Charnay et al. (2021) included the additional microphysical processes of nucleation on cloud condensation nuclei (CCNs), condensation, and coalescence, but only for the case of K2-18b. In addition, the particle size in their model was controlled by the number density of CCNs, a free parameter. As the particle size is a strong control of the optical properties of clouds, a more rigorous treatment of microphysical processes is needed.

In this work, we expand on these water-cloud models by taking into account the kinetics of microphysical processes and a variety of different nucleation pathways for a grid of Y-dwarf and temperate giant-planet atmospheres. We then provide insights for future observations of any exoplanets with H/He-dominated atmospheres and temperatures allowing for $H_2O$ condensation with simulated thermal emission and geometric albedo spectra. In Section 2, we describe the atmospheric models we use and the parameters we consider in our grid. In Section 3, we describe the impacts of microphysical processes on the water-cloud profile and compare them to those produced by a new version of the Ackerman & Marley (2001) model—`Virga`. We also generate simulated thermal emission and geometric albedo spectra to inform future observations. In Section 4, we discuss the implications of our results. We summarize our conclusions in Section 5.

## 2. Methods

### 2.1. Cloud Models

We use the 1D Community Aerosol and Radiation Model for Atmospheres (CARMA; Turco et al. 1979; Toon et al. 1988; Jacobson et al. 1994; Ackerman et al. 1995), to calculate the size and vertical distributions of water-cloud particles in temperate H/He atmospheres. CARMA is a general microphysical model that takes into account the nucleation, condensation, evaporation, coalescence/coagulation, sedimentation, and mixing of cloud particles of a variety of compositions in various atmospheres (Gao et al. 2018). We consider homogeneously nucleated water ice and heterogeneously nucleated water-ice particles, where the heterogeneously nucleated water-ice particles can nucleate on upwelled KCl cloud particles in both the Y-dwarf and temperate giant-planet cases. We also allow for water ice to heterogeneously nucleate on sedimenting meteoritic dust and photochemical hazes in the temperate giant-planet cases. We assume the water-ice particles are all spherical for simplicity.

We compare the CARMA cloud distributions to those computed by `Virga` to evaluate the importance of the microphysics of water-cloud formation. `Virga` (Batalha 2020) is an updated version of the Ackerman & Marley (2001) model, which is widely used in calculating cloud distributions in exoplanet and brown dwarf atmospheres (e.g., Cushing et al. 2008; Stephens et al. 2009; Morley et al. 2015; Rajan et al. 2017). `Virga` takes into consideration the sedimentation and mixing of cloud particles, with the vertical extent of the cloud controlled by the sedimentation efficiency parameter $f_{sed}$. Thus, comparing the two models will give us a measure of the effect of nucleation, condensation, and coalescence/coagulation on the cloud distribution. The comparison is done through the cloud optical depths, where we sweep through a range of $f_{sed}$ values to see which comes the closest to the CARMA profile. The background atmospheres, condensate properties, and particle-size grids are identical between the two models. For both CARMA and `Virga` we use Mie theory to calculate the cloud optical properties, including the optical depth, single-scattering albedo, and asymmetry parameter.

### 2.2. Model Grid

We compute the pressure–temperature (P-T) profiles for a grid of exoplanets and brown dwarfs assuming cloud-free background atmospheres and solar metallicity using a widely used thermal structure code (Saumon & Marley 2008; Morley et al. 2014b). Figure 1 shows the P-T and eddy diffusion coefficient ($K_{zz}$) profiles for both grids. The eddy diffusion coefficient is a parameterization of the vertical mixing in the atmosphere. In the convective region it is determined through mixing-length theory, while in the radiative region a minimum mixing length of a tenth of a scale height is assumed (Ackerman & Marley 2001; Morley et al. 2015). The Y-dwarf grid spans effective temperatures between 300 and 450 K and log($g$) between 3.5 and 4.5, with g in cgs units, based on Y-dwarf evolution models (Saumon & Marley 2008) and previous water-cloud models and retrievals (Morley et al. 2014b; Zalesky et al. 2019). The temperate giant-planet grid considers a Jupiter analog (surface gravity of 24.79 m s$^{-2}$) placed at 1, 2, and 3 au from a Sun-like host star. We run our cloud models using these model atmospheres. Cloud feedback on the P-T profiles and resulting spectra are not considered here as our goal is to compare the cloud morphology between the different models; while `Virga` can be coupled to the thermal structure code, CARMA cannot, and so we ignore cloud feedback. We discuss the impact of this in Section 4.

`Virga` does not consider the nucleation pathway of clouds. For the CARMA models, we examine homogeneous nucleation and heterogeneous nucleation with three types of CCNs. For both Y dwarfs and temperate Jupiters, we investigate the efficacy of KCl cloud particles mixed upwards from lower altitudes (Figure 1; Lodders 1999; Morley et al. 2012). The KCl clouds are simulated with CARMA assuming homogeneous nucleation from KCl vapor, as in Gao & Benneke (2018). In addition, for the temperate Jupiters, we also assess the role of photochemical hazes (West et al. 2009, p. 79; 2009, p. 161; Lavvas et al. 2011) and meteoritic dust (Draine 2004, p. 317; Moses & Poppe 2017), which we set as downward fluxes of particles from the top of the atmosphere (Lavvas & Koskinen 2017). The fluxes of CCN span from $10^{-12}$ to $10^{-20}$ g cm$^{-2}$ s$^{-1}$, which roughly covers the fluxes of the aforementioned particles (Lavvas & Koskinen 2017; Moses & Poppe 2017). Based on previous laboratory measurements (Israelachvili 2011; Yu et al. 2021), we parameterize the contact angles between water ice and photochemical hazes to be 60° and the contact angle between water ice and meteorite dust to be 0.1°. Meteorite dust has better wettability to water ice due to its high surface energy (Kloubek 1974; Israelachvili 2011). By considering sources of CCNs from above and below for the temperate Jupiters, we can evaluate how different CCNs impact the water-cloud distribution. All CCNs are allowed to grow through coagulation.

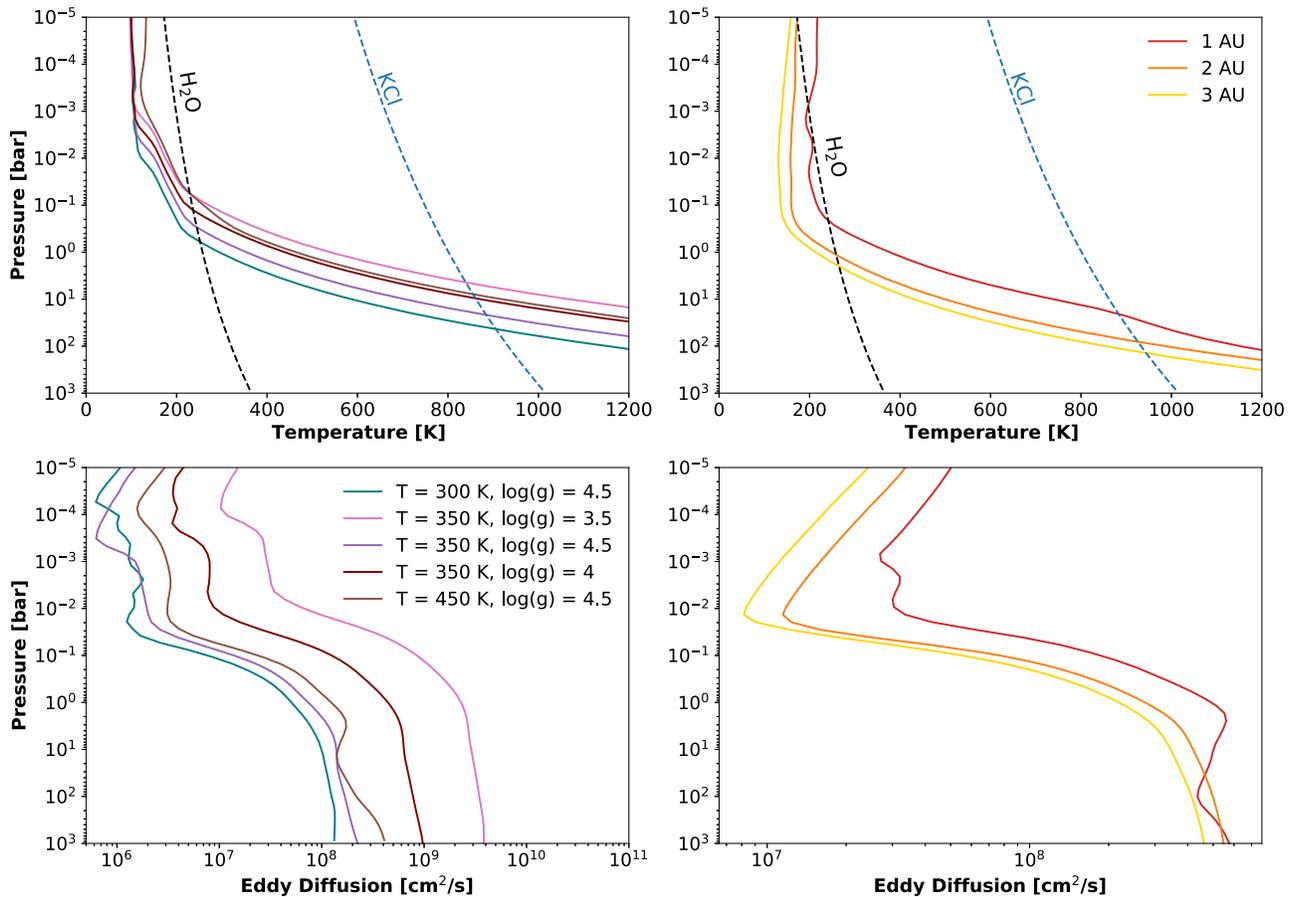

**Figure 1.** Pressure–temperature (top) and eddy diffusion coefficient (bottom) profiles of our modeled Y dwarfs (left) and temperate giant planets (right). Condensation curves of KCl (blue) and water ice (black) are shown for comparison.

For both CARMA and `Virga`, we assume a deep water vapor mixing ratio of ~$8 \times 10^{-4}$. The exact value assumes elemental oxygen depletion due to sequestration in deep forsterite clouds, with all remaining O being in water vapor. For CARMA, we assume a cloud-free atmosphere as an initial condition, with water vapor at the bottom of the atmosphere fixed to the aforementioned mixing ratio. As the model runs, the water vapor is mixed upwards until saturation is reached, after which it nucleates into pure or heterogeneous water-ice particles that can then grow via condensation, be lost to evaporation, and get transported via sedimentation and mixing. We then compare the steady-state CARMA cloud optical depth profiles to the same profiles computed by `Virga` for a suite of $f_{sed}$ values.

### 2.3. Spectral Generation

In preparation for future observations and to evaluate the impact of the differences between CARMA and `Virga` on observables, we generate thermal emission spectra for the Y-dwarf cases and optical geometric albedo spectra for the Jupiter-analog cases. We use `PICASO` (Batalha et al. 2019) to generate the geometric albedo spectra, which span 0.3–1.0 $\mu$m. The spectra include the opacities from five chemical species: $H_2O$, $CH_4$, $NH_3$, $H_2$, K, and Na.

We generate line-by-line thermal emission spectra for the Y-dwarf models using the code described in the Appendix of Morley et al. (2015), which includes absorption and scattering from molecules, atoms, and clouds. The spectra are generated for an evenly spaced wavenumber grid from 40 to 33333 cm$^{-1}$ in 1 cm$^{-1}$ steps. We include opacities from 17 species: $H_2$, He, CO, $CO_2$, $H_2O$, $CH_4$, $NH_3$, $PH_3$, $H_2S$, Na, K, TiO, VO, FeH, CrH, Rb, and Cs, as well as collision-induced opacity of $H_2$–$H_2$, $H_2$–He, $H_2$–H, and $H_2$–$CH_4$ using Richard et al. (2012). The model opacity database for both thermal emission and reflected light is based on Freedman et al. (2008), with significant updates as described in Freedman et al. (2014).

### 3. Results

In the following sections, we present the optical depth profiles for the water-ice clouds in both of our grids as generated by CARMA in comparison to `Virga`. For Y dwarfs, we focus on the 4.074 $\mu$m slice for each respective profile, while for temperate giant planets we choose 0.642 $\mu$m. There is little wavelength dependence in the optical depth profiles due to the large size of the cloud particles at the probed pressure levels; these wavelengths were selected based on where we expect observations of these objects to be the brightest (Skemer et al. 2014; MacDonald et al. 2018). We then go on to predict the geometric albedo and thermal emission spectra of these bodies to aid future observations.

### 3.1. Y-dwarf Water-cloud Profiles

No water cloud computed by `Virga` fully reproduces the entire CARMA cloud (Figure 2, left). Currently, water-cloud models under the Ackerman & Marley framework generally use $f_{sed}$ values of 3–8 (Morley et al. 2014b, 2018) and `Virga` models within the range of 3–6 do present themselves to be the closest to the CARMA cloud near the cloud base in the heterogeneous case. We also find the CARMA cloud transitions from a higher $f_{sed}$ value at the cloud top toward a lower $f_{sed}$ at the cloud base for the heterogeneous cases. In contrast, in the homogeneous case, the CARMA water cloud transitions to a lower $f_{sed}$ value at the top of the cloud but is closest to the near-cloudless $f_{sed} = 10$ `Virga` case at the cloud base. The homogeneously nucleated water cloud is more vertically extended than the heterogeneously nucleated water cloud due to the fact that the latter nucleates on KCl particles mixed upwards from below, and as such the water-cloud profile follows the KCl cloud profile, which has a smaller scale height due to KCl's higher mass density. The `Virga` clouds are more optically thick, up to two orders of magnitude higher at the cloud base for the homogeneously nucleated case. The CARMA clouds on the other hand are more vertically extended as more nucleation occurs at the top of the cloud. The water vapor is supersaturated above 1 mbar due to eddy mixing that drives the nucleation at the top.

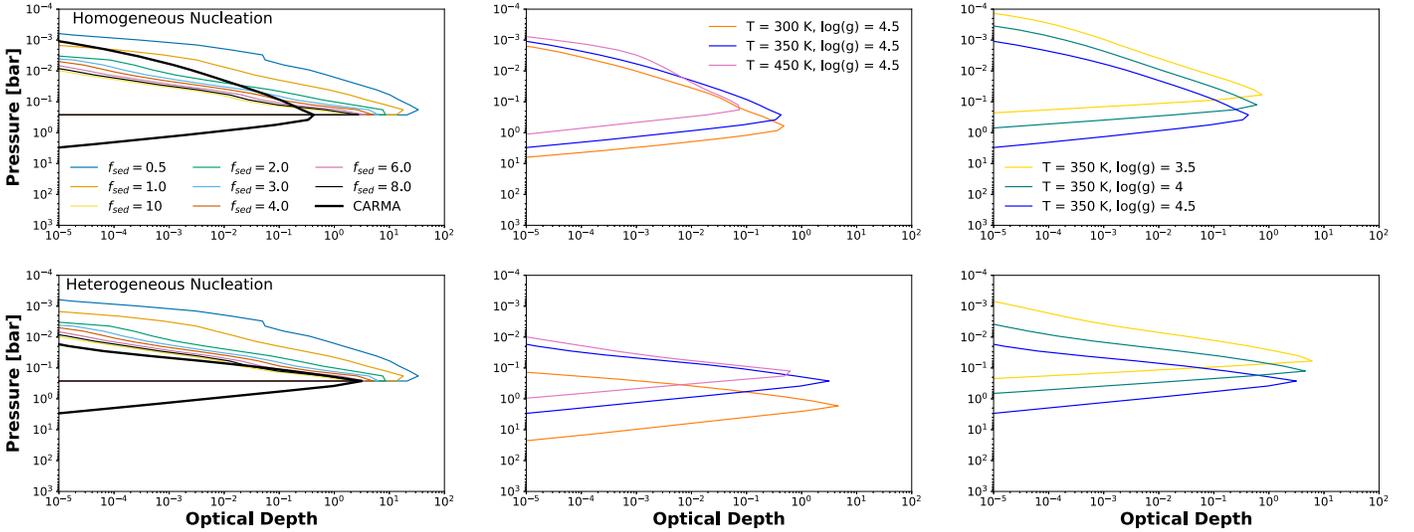

**Figure 2.** Homogeneously (top) and heterogeneously (bottom) nucleated water-ice cloud optical depth profiles from our Y-dwarf models. Left: comparison of CARMA and select `Virga` cloud profiles with varying $f_{sed}$ values for the $T_{eff} = 350$ K, $\log(g) = 4.5$ case. Middle: CARMA cloud variations in optical depth profiles due to differences in $T_{eff}$. Right: CARMA cloud variations in optical depth profiles due to differences in $\log(g)$.

Increasing $T_{eff}$ while fixing gravity results in the cloud profile shifting upwards to lower pressure levels due to the higher local temperatures. As the available condensate mass also decreases with decreasing pressure, the cloud optical depth also decreases. In addition, at sufficiently high temperatures (e.g., 450 K), nucleation becomes inefficient, further decreasing cloud optical depth. Increasing gravity at a fixed $T_{eff}$ results in lower atmospheric temperatures due to lower atmospheric gas opacity, leading to deeper cloud bases. The lower condensate mass and higher sedimentation velocities at higher gravity then lead to decreased cloud optical depths.

The single-scattering albedo of the CARMA and `Virga` clouds are both close to 1.0 across the entire cloud profile. The asymmetry parameter of the CARMA cloud is about 2% larger than the `Virga` clouds due to larger particles in the CARMA models stemming from rapid particle growth, which are more forward scattering. The particle sizes vary across the vertical layers of the atmosphere, from 10 to 100 $\mu$m near the cloud base in the lower atmosphere ($1–10^{-2}$ bar) to submicron sizes near the cloud top ($10^{-2}–10^{-6}$ bar).

### 3.2. Y-dwarf Thermal Emission Spectra

We generate Y-dwarf thermal emission spectra (Figure 3), which show distinct absorption features that can be attributed to $H_2O$, $CH_4$, $NH_3$, CO, and $CO_2$ (Morley et al. 2014b). No `Virga` case with $f_{sed} \leqslant 10$ matches the homogeneous CARMA spectrum due to the low optical depth of the homogeneously nucleated clouds, though they are still sufficiently optically thick to decrease the emission, particularly in the 4–5 $\mu$m region, when compared to the cloudless case. Meanwhile, the heterogeneous CARMA spectrum can best be matched by `Virga` models with $f_{sed}$ between 4 and 6, in line with previous fits of the Ackerman & Marley model to Y-dwarf observations (Morley et al. 2018). These "best-fit" $f_{sed}$ values vary with the temperature and gravity of the Y dwarf. The difference between the CARMA and `Virga` thermal emission spectra varies with wavelength and can be attributed to the different vertical profiles of the optical properties of the two model clouds, which in turn allow for emission to probe different parts of the atmosphere and gas abundances. We find that the largest differences between the spectra occur at absorption bands of $NH_3$, CO, and $CO_2$. Therefore, knowing the vertical opacity profile of the water clouds is essential to accurately estimate the abundances of these gases.

To further examine the CARMA and `Virga` spectra across the model grid, Figure 4 shows thermal emission spectra in the $Y$, $J$, $H$, and $M$ bands where prominent emission is seen. The `Virga` cases with $f_{sed} = 3$, across both grids and all bands, have lower flux levels in their thermal emission spectra due to the higher optical depth of the `Virga` clouds compared to CARMA. Higher gravity cases lead to higher fluxes because the clouds are lower in the atmosphere, allowing for a deeper probe into the atmospheres of these Y dwarfs. The impact of different cloud models is particularly pronounced in the 4–5 $\mu$m region ($M$ band), where Y dwarfs are the brightest and most amenable to characterization by, e.g., JWST. For example, the `Virga` case with $T = 300$ K and $f_{sed} = 3$ does not show the same spectral features as those at other wavelengths, which could be due to the fact that the `Virga` clouds are about $10^2$ times optically thicker than the CARMA clouds, therefore quenching the flux in this region. In general, there are multiple orders of magnitude differences between the CARMA and `Virga` thermal emission fluxes.

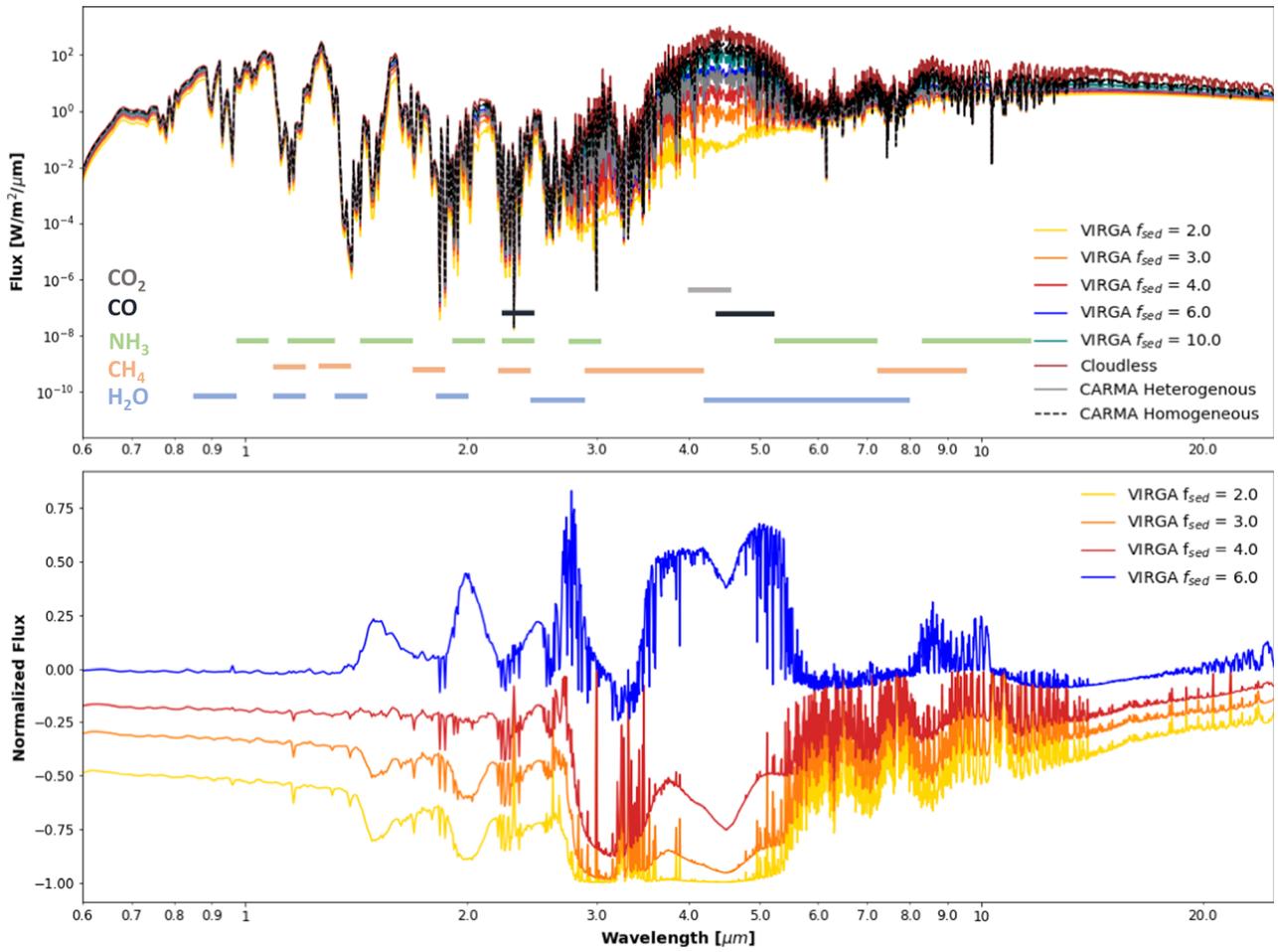

**Figure 3.** Top: thermal emission spectra of a cloudless atmosphere, various `Virga` models, and both homogeneously and heterogeneously nucleated water-ice cloud CARMA cases for a Y dwarf with $T_{\text{eff}} = 350$ K, $\log(g) = 4.5$. The molecular absorption bands are labeled. Bottom: fractional difference between the heterogeneously nucleated CARMA case and select `Virga` cases.

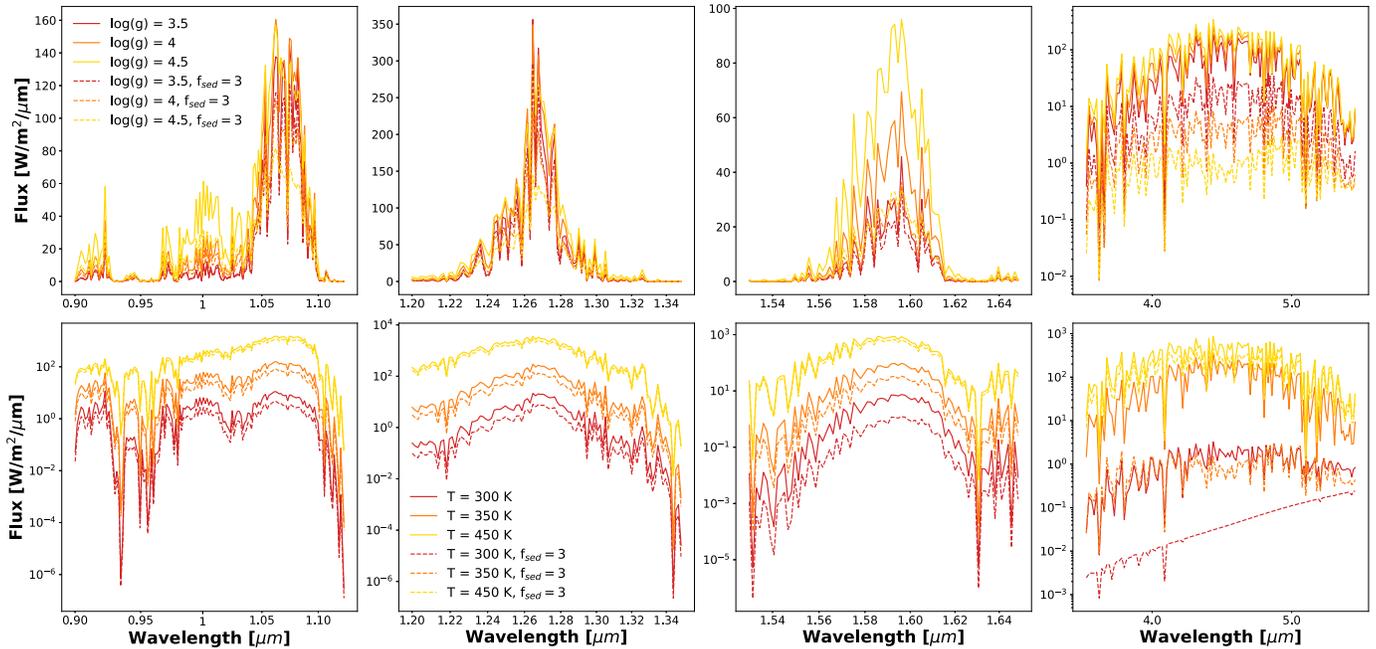

**Figure 4.** Simulated thermal emission spectra of Y dwarfs in the $Y$, $J$, $H$, and mid-infrared bands for the heterogeneously nucleated water-ice cloud CARMA model (solid) and a `Virga` model with $f_{\text{sed}} = 3$ (dashed). Top: variations in thermal emission spectra with $\log(g)$ at $T_{\text{eff}} = 350$ K. Bottom: variations in thermal emission spectra with $T_{\text{eff}}$ at $\log(g) = 4.5$.

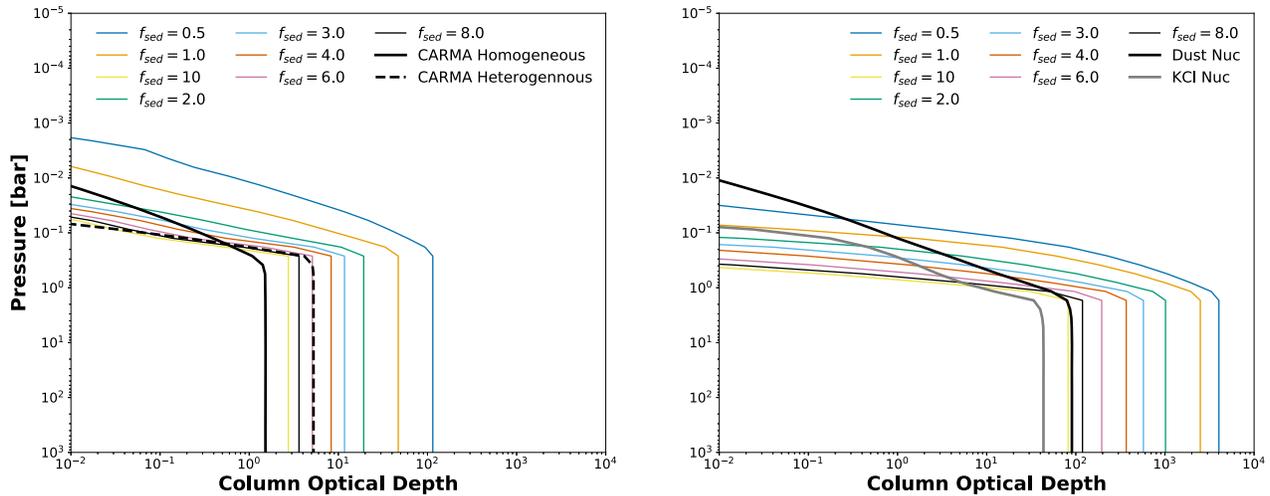

**Figure 5.** Cumulative optical depth profiles for (left) a Y dwarf with $T_{eff} = 350$ K, $\log(g) = 4.5$, and (right) a temperate giant planet at 3 au, CCN flux $= 10^{-14}$.

The cumulative optical depth cloud profile can help us understand the impact of clouds on observations by showing where the cloud optical depth ∼1 and the slope of the profile near that location (Figure 5). We find that the CARMA heterogeneous cloud matches closest to `Virga` clouds with $f_{sed} = 6$ and 8 at optical depth unity, while the CARMA homogeneous water cloud matches $f_{sed} = 10$ at the same point. The CARMA thermal emission spectra are less bright than what we expect given the cumulative optical depth, matching more closely models with $f_{sed} < 6$. This discrepancy is likely due to the rapid increase in opacity of the CARMA cloud below the optical depth unity point such that at the cloud base it is slightly more optically thick than the $f_{sed} = 6$ `Virga` cloud.

### 3.3. Temperate Giant-planet Cloud Profiles

No significant amount of homogeneously nucleated water clouds form in the temperate giant-planet atmospheres. This is because their temperature–pressure profiles are almost isothermal in our atmospheric region of interest between 1 bar and 1 $\mu$bar (Figure 1), which decreases the water supersaturation and consequently, reduces the rates of homogeneous nucleation. We, therefore, focus only on water clouds generated by heterogeneous nucleation.

The CARMA clouds that nucleated on photochemical hazes and meteoritic dust are much more vertically extended and much less optically thick in comparison to the `Virga` clouds (Figure 6). Because hazes and dust rain downwards into the atmosphere, they result in greater rates of nucleation at the top of the cloud in our CARMA model. The cloud optical depth increases as the planet is placed farther away from the host star as this decreases the equilibrium temperature of the planet. Interestingly, there is a larger difference between the 1 and 2 au cases compared to the 2 and 3 au cases. This is due to the decreased nucleation efficiency in the 1 au case. Also, at 2 and 3 au, the cloud base lies in the convective region of the atmosphere. Therefore, the internal flux of the planet becomes more important than the solar flux when determining the location of the cloud base. As with the Y-dwarf models, the water clouds that nucleated on the KCl clouds are more vertically constrained than the water clouds that nucleated on the hazes/dust. Increasing the contact angle reduces the nucleation rate and therefore the optical depth, and vice versa.

Interestingly, the cloud optical depth does not vary significantly with downward CCN flux until it is $<10^{-18}$ g cm$^{-2}$ s$^{-1}$ (Figures 6 and 7). This is due to the fact that, at higher fluxes, the CCN number density is limited by rapid coagulation, which in turn sets a limit on the heterogeneous nucleation rate because it is linearly dependent on the CCN number density. This is in contrast to the results of Gao et al. (2018), which showed a linear relationship between cloud optical depth and downward CCN flux, but in that work, CCN coagulation was ignored.

### 3.4. Temperate Giant-planet Geometric Albedo Spectra

The geometric albedo spectra of the temperate giant planets as simulated with PICASO can be seen in Figure 8, which also includes the absorption features that can be attributed to $H_2O$, $CH_4$, and $NH_3$ (Cahoy et al. 2010; MacDonald et al. 2018). We ignore the contribution to the geometric albedo from the CCNs because they have negligible optical depth compared to the water-ice cloud. No `Virga` case can duplicate the wavelength dependency of the CARMA cases. Specifically, the continuum regions of the CARMA case have a lower albedo than those of the `Virga` cases, while the absorption features are relatively shallower. These regions correspond to $CH_4$ absorption. The CARMA clouds have a lower geometric albedo overall because they are optically thinner, allowing for deeper probes into the atmosphere and therefore more light absorption. They are also more vertically extended than the `Virga` clouds, which cause some attenuation of the absorption features between 1 and $10^{-2}$ bars, where the CARMA cloud has a much shallower drop-off in optical depth compared to the `Virga` models. When KCl acts as the CCN, the clouds are optically thinner still and therefore less reflective compared to the water clouds with hazes and dust acting as CCN.

We find a large discrepancy between the geometric albedo spectra and what is expected from the cumulative optical depth profiles (Figure 5). The CARMA heterogeneous water cloud that nucleated on KCl matches the `Virga` $f_{sed} = 3$ cloud, while the dust nucleation case is closest to $f_{sed} = 1$ at optical depth unity. In other words, the `Virga` water clouds tend to be much brighter than expected. This difference is a result of the much more rapid increase in the opacity of the `Virga` clouds with decreasing altitude.

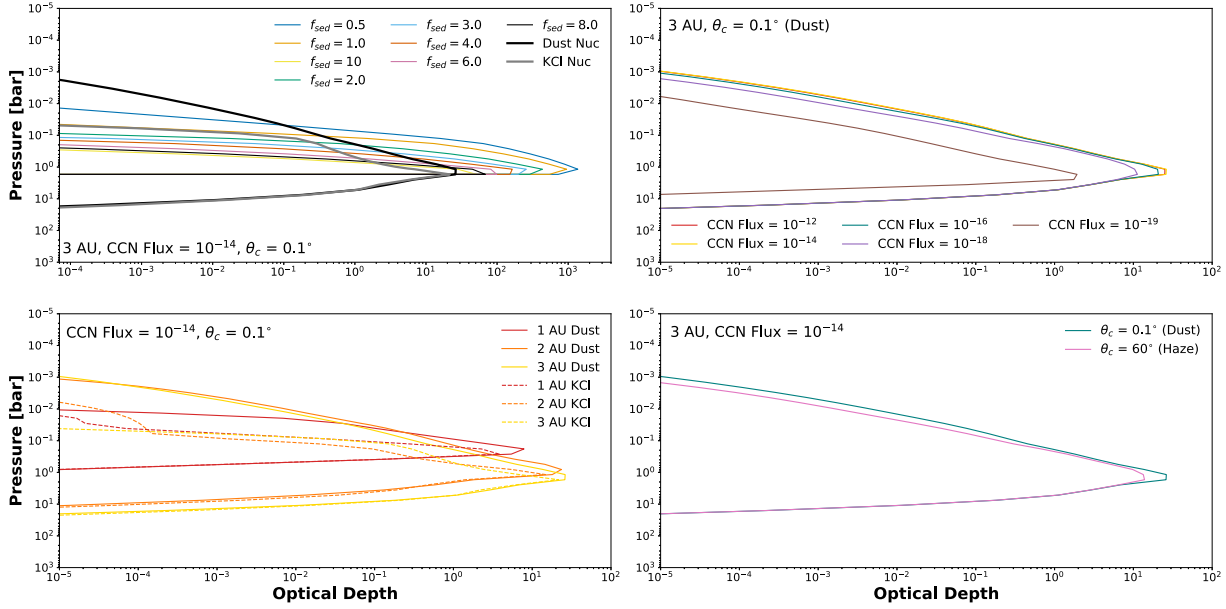

**Figure 6.** Top left: comparison of the water-ice cloud optical depth profiles of the heterogeneously nucleated CARMA cases and select `Virga` cases with varying $f_{sed}$ values for the 3 au temperate giant-planet model. The other panels show the variations in the optical depth profiles of the heterogeneously nucleated water-ice clouds with (top right) varying downward CCN flux, (bottom left) planet semimajor axis and CCN source, and (bottom right) contact angle, assuming dust/haze CCNs. CCN flux is in units of g cm$^{-2}$ s$^{-1}$.

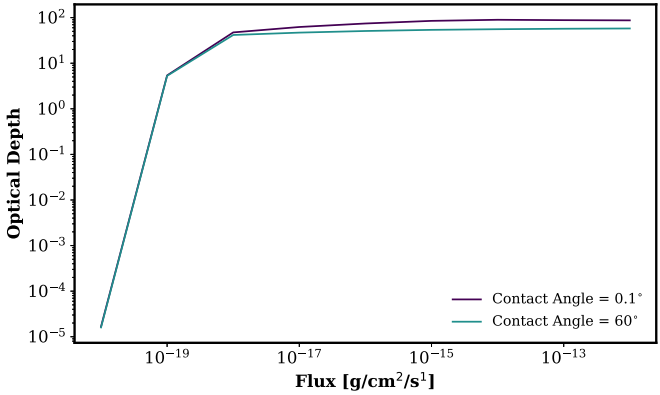

**Figure 7.** The total water-cloud optical depth of a temperate giant planet at 3 au as a function of downward CCN flux for contact angles of 0.1° (purple) and 60° (turquoise).

When comparing the spectra across the 1, 2, and 3 au temperate giant-planet grid, we find that the 2 au case is more reflective than the 3 au case beyond 0.7 $\mu$m, as the water cloud on the 3 au planet is deeper in the atmosphere. The 1 au case does not have as optically thick of a cloud as the 2 and 3 au cases, causing it to have a lower geometric albedo overall. Higher-contact-angle clouds, which correspond to hazes, are optically thinner, leading to lower albedo. The progression in geometric albedo spectra with decreasing CCN flux matches our findings that the cloud optical depth does not significantly decrease with decreasing CCN flux appreciably until the flux is

$<10^{-18}$ g cm$^{-2}$ s$^{-1}$. Below this flux, the geometric albedo spectra rapidly converge to the cloudless case.

## 4. Discussion

Consideration of the microphysics of water-cloud formation on cool substellar objects shows that the optical properties and distributions of these clouds are sensitive to the nucleation energy barrier and pathway and the source and makeup of the CCNs. Homogeneous nucleation produces vertically extended, optically thin clouds due to low nucleation rates at the cloud base and enhanced nucleation rates at higher (colder) altitudes. Heterogeneous nucleation on CCN sourced from the upper atmosphere produces vertically extended clouds while those sourced from CCNs mixed upwards from below produces vertically constrained clouds, though this may be due to our assumed CCN's higher mass density. Comparisons between CARMA and `Virga` show that it is difficult for the latter model to reproduce the lower cloud base optical depths produced by CARMA resulting from the nucleation barrier, though previously used $f_{sed}$ values of 3–6 do match the CARMA results reasonably closely for the heterogeneous nucleation cases.

The differences in cloud distributions between CARMA and `Virga` translate to wavelength-dependent differences in thermal emission and geometric albedo spectra, in particular the depth of gas absorption bands. Investigations of Y dwarfs in the 4–5 $\mu$m region by space and ground-based observatories will be especially sensitive to cloud assumptions. Future mock retrievals on spectra produced from different cloud models should be attempted to ascertain the impact different water-cloud treatments have on the retrieved gas abundances in cooler worlds (e.g., Mukherjee et al. 2021).

We do not include radiative cloud feedback in our study in order to focus on the morphology of the cloud itself and emphasize the importance of microphysical processes. If included, the additional cloud opacity would increase the temperature below the cloud by trapping more of the internal flux, which in turn would affect the ability of water to condense, the cloud optical depth, and local convective stability (Morley et al. 2014b). On the other hand, for temperate giant planets, the reflectivity of water-ice clouds would reduce the flux absorbed by the planet and thus the temperature, and therefore how the P-T profiles of these objects respond to the formation of water clouds is more complex. In addition, we also do not consider patchy clouds, which are characteristic of water clouds observed on Earth, Jupiter, and Saturn. Horizontal heterogeneity in the cloud cover would lead to a superposition of clear and cloudy spectra and also temporal variability (Marley et al. 2010; Morley et al. 2014a). This would therefore result in the sensitivity of the gas spectral features to both the vertical and horizontal cloud distributions.

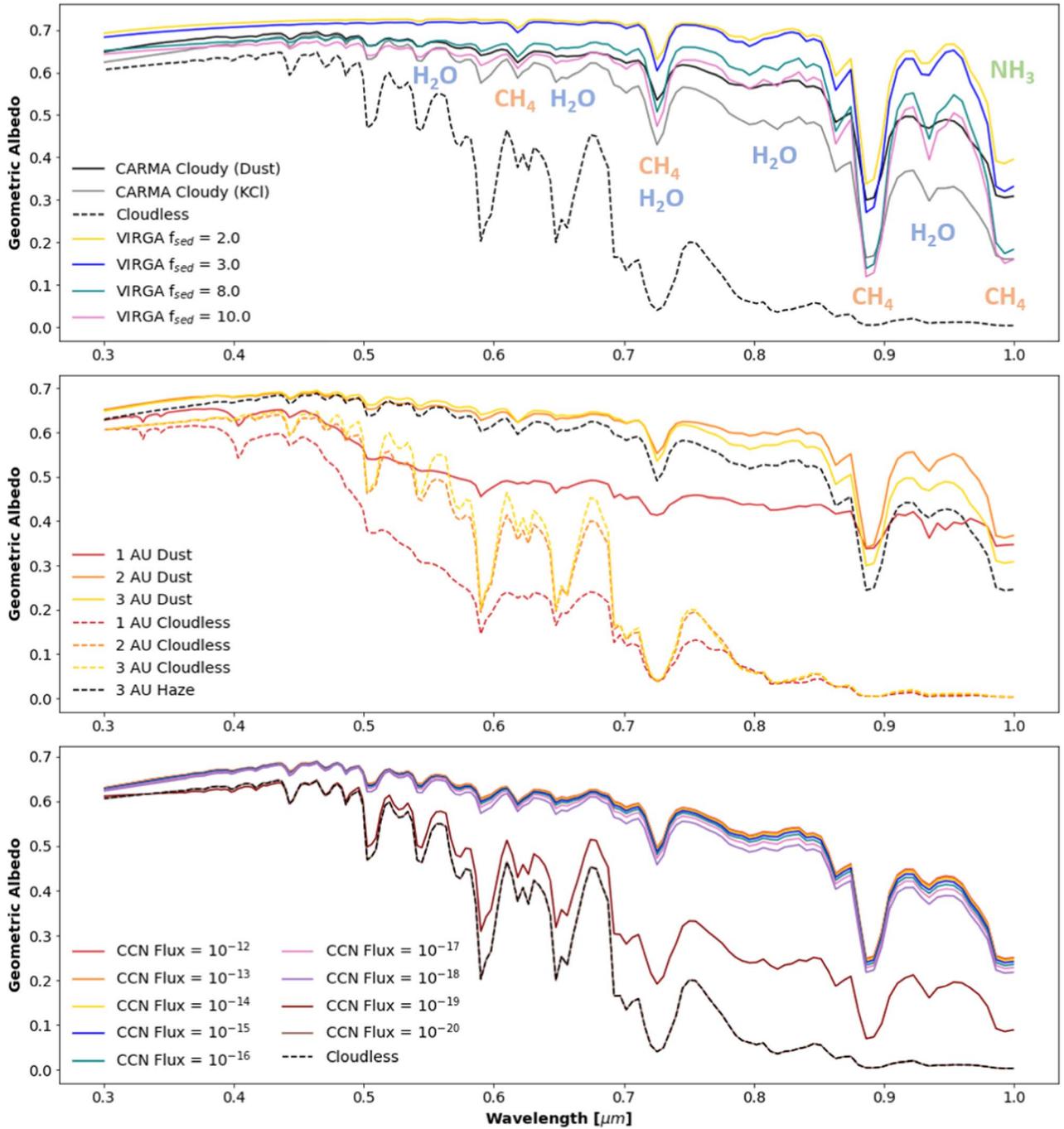

**Figure 8.** Simulated geometric albedo spectra of the temperate giant-planet grid generated with PICASO. Top: comparison of the heterogeneously nucleated water-cloud CARMA cases, a cloudless case, and select `Virga` cases for a 3 au model. The molecular absorption bands are labeled. Middle: variations of the geometric albedo spectra with varying contact angle (dust vs. haze) and planet semimajor axis, compared to corresponding cloudless spectra. Bottom: variations of the geometric albedo spectra with varying downward CCN flux for the 3 au model. The flux = $10^{-20}$ g cm$^{-2}$ s$^{-1}$ case is consistent with being cloudless. CCN flux is in units of g cm$^{-2}$ s$^{-1}$.

We also ignore the mass-loading effect and latent heat release in our model and their impact on the local atmosphere. The latent heat release from water condensation would reduce the temperature gradient at and above the altitude of the water-cloud base, leading to lower rates of water-ice nucleation, but the effect may only be pronounced for-higher metallicity ($\geqslant 10\times$ solar) objects (Tang et al. 2021). The mass-loading effect, which could lead to periodic storms in H/He atmospheres owing to the higher molecular weight of water, would result in the greater temporal and spatial variability of water clouds in Y-dwarf and temperate giant-exoplanet atmospheres (Cushing et al. 2016). The periodicity and intensity of these storms would depend on the metallicity and could be reduced for solar-metallicity atmospheres (Li & Ingersoll 2015). Physical properties of the H/He atmosphere in addition to molecular weight, like the atmospheric viscosity, thermal conductivity, and heat capacity, and the collisional diameter of atmospheric molecules should have little impact on the vertical and size distributions of water-cloud particles (Gao & Benneke 2018).

Water-ice cloud particles need not be spheres, as assumed in our model, and in fact could form plates, needles, and other crystalline shapes (Murray et al. 2015). We expect the nonspherical nature of water-ice particles to lower the asymmetry parameter and thus reduce forward scattering (Kuo et al. 2016).

## 5. Conclusions

Future observations of Y-type brown dwarfs and temperate giant planets will be affected by the water clouds present in their atmospheres. Here we studied the morphology and microphysics of water-ice clouds of cool objects by comparing the results of two cloud models, CARMA and Virga, under different microphysical parameter values. Our main findings are the following:

1. No Virga clouds fully recreate the CARMA clouds. The homogeneously nucleated CARMA clouds on Y dwarfs are similar to Virga clouds with $f_{sed} = 10$ at the cloud base but are more vertically extended than the Virga clouds. CARMA clouds that heterogeneously nucleated on upwelled KCl cloud particles are closer to Virga clouds with $f_{sed} = 3$–6, a value used in previous studies that interpreted Y-dwarf observations.

2. Thermal emission spectra predicted by CARMA and Virga for the Y dwarfs show the largest differences at 4–5 $\mu m$, with smaller wavelength-dependent differences throughout the rest of the spectra corresponding to absorption bands of $H_2O$, $CH_4$, $NH_3$, CO, and $CO_2$. This reflects the sensitivity of gas species features to the cloud vertical distribution and has important implications for future attempts to characterize Y dwarfs in the $M$ band.

3. There is no significant homogeneous nucleation of water clouds in temperate giant-planet atmospheres. For water clouds heterogeneously nucleating on CCNs sourced from the upper atmosphere, e.g., photochemical haze and meteoritic dust, the cloud optical depth does not vary strongly with downward CCN flux until it falls below $10^{-18}$ g cm$^{-2}$ s$^{-1}$ due to CCN coagulation.

4. CARMA clouds generally produce lower geometric albedos on temperate giant planets because they are optically thinner than the equivalent Virga models. However, the gas absorption bands are brighter due to the CARMA clouds being more vertically extended.

Y dwarfs and temperate giant exoplanets serve as a link between the hotter exoplanets and brown dwarfs we have characterized thus far and the solar system gas giants. Understanding this link will require more physical treatments of water clouds in models of their atmospheres and future observational characterization. Broad wavelength observations of Y-dwarf thermal emission spectra by JWST could help constrain the cloud vertical extent and the gas abundance profiles. The same could be done for temperate giant exoplanets with the reflected light spectra collected by the Nancy Grace Roman Space Telescope. Interpreting these data would benefit from established analysis methods for the solar system giant planets (e.g., Sromovsky et al. 2019). At the same time, models of cooler atmospheres must take into account the ways in which water clouds impact H/He atmospheres by including cloud microphysics, radiative feedback, the mass-loading effect, latent heat release, etc., in order to more rigorously interpret the upcoming data.

We thank J. C. Y. Mang, W. Mang, and E. Chan for their continual support throughout this project. P.G. acknowledges support from the 51 Pegasi b Fellowship funded by the Heising-Simons Foundation, and NASA through the NASA Hubble Fellowship grant No. HST-HF2-51456.001-A awarded by the Space Telescope Science Institute, which is operated by the Association of Universities for Research in Astronomy, Inc., for NASA, under contract NAS5-26555.

*Software:* CARMA (Ackerman et al. 1995), PICASO (Batalha et al. 2021), Virga (Batalha 2020), Jupyter (Kluyver et al. 2016), NumPy (Walt et al. 2011), SciPy (Jones et al. 2001), Matplotlib (Hunter 2007).

## ORCID iDs

James Mang ● https://orcid.org/0000-0001-5864-9599
Peter Gao ● https://orcid.org/0000-0002-8518-9601
Callie E. Hood ● https://orcid.org/0000-0003-1150-7889
Jonathan J. Fortney ● https://orcid.org/0000-0002-9843-4354
Natasha Batalha ● https://orcid.org/0000-0003-1240-6844
Xinting Yu ● https://orcid.org/0000-0002-7479-1437
Imke de Pater ● https://orcid.org/0000-0002-4278-3168